\documentclass[%
notitlepage,
twocolumn,
nolongbibliography,
 amsmath,amssymb,
prb,
]{revtex4-2}

\usepackage{graphicx}% Include figure files
\usepackage{dcolumn}% Align table columns on decimal point
\usepackage{bm}% bold math
\usepackage{siunitx}
\usepackage{ulem}		% to use \sout command

\usepackage{xcolor}
\usepackage{mathtools}
\usepackage{bbold}
\usepackage[]{graphicx}
\usepackage{tabularx}
\usepackage[hidelinks]{hyperref}

\hypersetup{colorlinks=true, citecolor=blue, urlcolor=black, linkcolor=blue}

\usepackage{empheq}

\usepackage[most]{tcolorbox}

\tcbset{colback=yellow!10!white, colframe=black!50!black, 
        highlight math style= {enhanced, %<-- needed for the ’remember’ options
            colframe=black,colback=red!10!white,boxsep=0pt}
        }

\newcommand{\QQ}{\mathbf{Q}}
\newcommand{\qq}{\mathbf{q}}

\definecolor{amber}{rgb}{0.82, 0.1, 0.26}

\newcommand{\changes}[1]{{\color[rgb]{0,0,0}{#1}}} % general comments

\begin{document}
%\Large{Key results: Förster-type interaction between molecule and TMDC}
\normalsize
%\tableofcontents
\preprint{APS/123-QED}

\title{Suppression and amplification of phonon sidebands in transition metal dichalcogenides \changes{by optical feedback}}

	\author{Thomas Tenzler}
	\email{t.tenzler@tu-berlin.de}
	\author{Andreas Knorr}
	\email{andreas.knorr@tu-berlin.de}
	\author{Manuel Katzer}
	\email{manuel.katzer@physik.tu-berlin.de}
    \affiliation{Technische Universit\"at Berlin, Institut f\"ur Theoretische Physik, Nichtlineare Optik und Quantenelektronik, Hardenbergstra{\ss}e 36, 10623 Berlin, Germany}

\date{\today}

\begin{abstract}
    Transition metal dichalcogenides (TMDCs) combine both strong light-matter-interaction and strong Coulomb-interaction for the formation of optically excitable excitons. Through radiative feedback control, a mechanism to control the linewidth can be applied, which modifies optical transition spectra. Here, we extend these investigations to the absorption spectra of TMDCs in a variety of geometries with respect to non-Markovian exciton-phonon-scattering contributions. Our approach is based on the self consistent solution of the microscopic Bloch equations and the macroscopic solution of the wave equation. \changes{We discuss the formation of a phonon sideband for MoSe$_2$ embedded in SiO$_2$, and two setups for enhancing or suppressing the phonon sideband in the spectrum.}

\end{abstract}

\maketitle
\section{Introduction}
Transition metal dichalcogenides (TMDCs) are two dimensional atomically thin semiconductors, which exhibit strong Coulomb-interaction leading to the formation of bound electron-hole pairs, called excitons, with high binding energies of hundreds of $\si{meV}$~\cite{MalteMoSelinewidth,haugkoch,DominikSelfenergy,Florian,katzer2023impact,he2014tightly,wang2018colloquium}. TMDC excitons also exhibit strong light-matter interaction, which is almost two orders of magnitude larger than in their bulk equivalent~\cite{13ReferenzKontrolle,MalteMoSelinewidth}. With these properties, TMDCs exhibit high values of their reflectivity~\cite{13ReferenzKontrolle} and are candidates for various applications in optoelectronics~\cite{13ReferenzKontrolle,Toulouse}. One possibility to control \changes{the function of} corresponding devices is coherent feedback of the radiative emission and provides the possibility to influence the linewidth and lineshifts of the optical spectra~\cite{Toulouse,13ReferenzKontrolle,Florian,Stroucken}. The mechanism is well studied with wide applications as for example in the feedback control of quantum well structures~\cite{finsterholzl2020nonequilibrium,barkemeyer2020revisiting}, the feedback control of the radiative lifetime~\cite{Toulouse} or as a control mechanism of the radiative linewidth~\cite{Florian,13ReferenzKontrolle} which can all influence the optical properties of the device in question. 

In this work we study half-sided cavities~\cite{Florian,carmele2013single,hein2014optical,rogers2020coherent,epstein2020near}, where we focus on the coherent control of the optical linewidth, which can be achieved by introducing and varying the position of the external mirror of the cavity with respect to the TMDC~\cite{Florian,13ReferenzKontrolle,rogers2020coherent,epstein2020near,zhou2020controlling,horng2019engineering} or varying the thickness of the encapsulation~\cite{Toulouse,epstein2020near}. This way the absorption depends on the time delay of the back coupled emission, suggesting that microelectronical mechanical devices can be used to achieve low energy cost and near perfect absorption~\cite{13ReferenzKontrolle, ansari2020nearly}.\\ 

\changes{In TMDCs, only} a fraction of excitons can interact optically, with most excitons having too much kinetic energy and are thus 'momentum-dark'~\cite{zhang2015experimental,MalteMoSelinewidth,katzer2023impact,selig2019ultrafast}. Their dynamics are activated by exciton-phonon scattering, leading to the formation of phonon sidebands, also where the transition energy is shifted by the polaron energy. This is theoretically explained by a non-Markovian approach~\cite{DominikSelfenergy}, \changes{which also inspired work of other groups~\cite{shree2018observation,FrankLengers}} and is expanded upon in this work by inclusion of feedback control. 
In particular, we discuss the possibility of the amplification or suppression of phonon sidebands in linear spectroscopy by coherent feedback control.
This is realized by a self consistent solution of the macroscopic wave equation, where the dipole source is determined by the microscopic semiconductor-Bloch equations.
The article is organized as follows: In Sec.~\ref{Sec:Theo}, we first introduce the interface/geometry, namely a MoSe$_2$ TMDC monolayer encapsulated in SiO$_2$ with a mirror positioned behind the encapsulation, which introduces the main mechanism of coherent feedback control. Afterwards we give details on the excitonic response and the dielectric environment. In Sec.~\ref{Sec:Disc} we discuss the theoretical results of the suppression or amplification of the phonon sidebands of the TMDC at room temperature by varying the mirrors position. 

\section{Theoretical Description}\label{Sec:Theo}
\subsection{Homogeneous environment}
\begin{figure}[t!]
    \centering
    \includegraphics[width=0.95\linewidth]{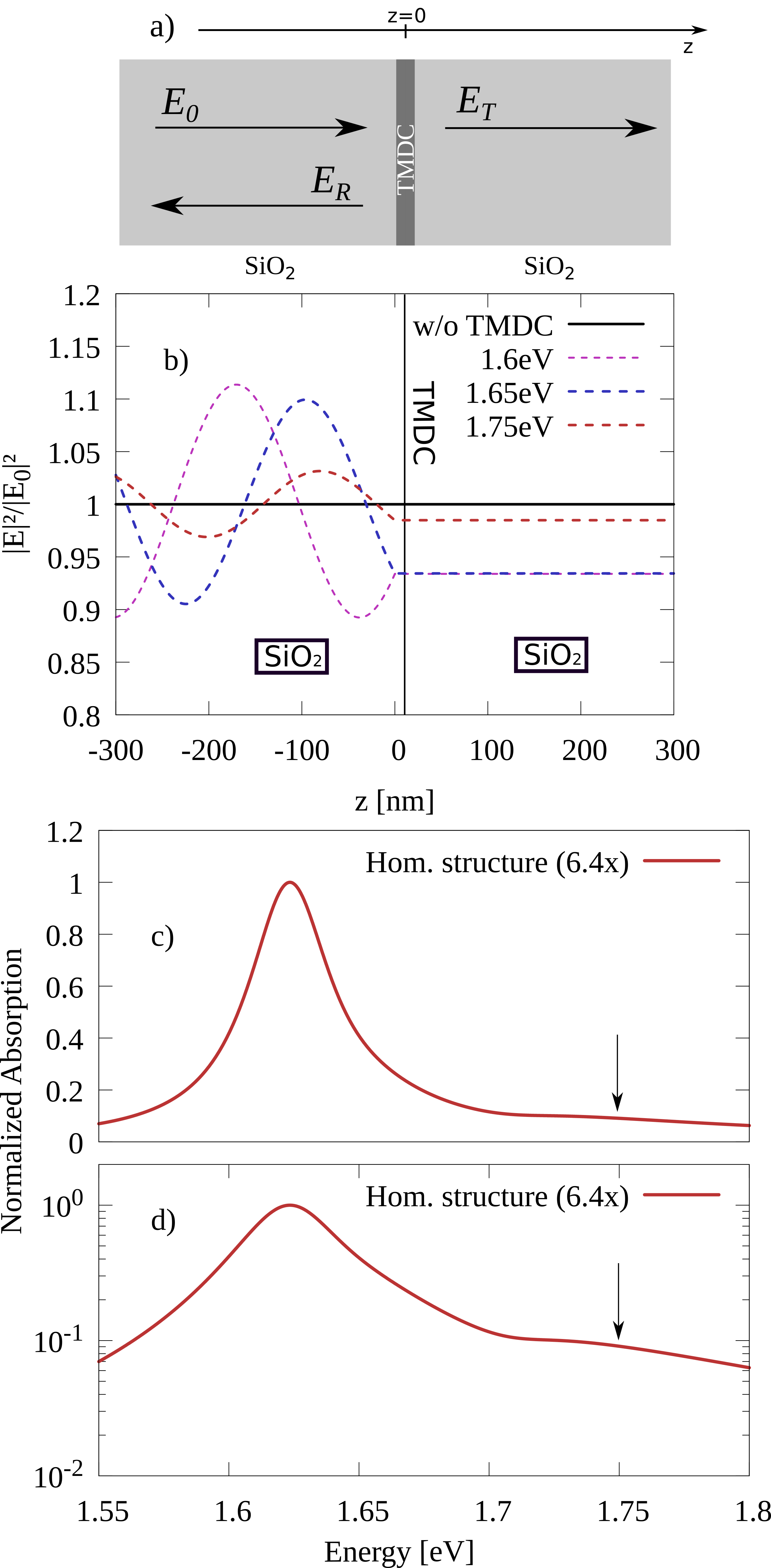}
    \caption{\changes{a) Sketch of the TMDC embedded in a homogenous SiO$_2$ environment and the incident field in front and behind the TMDC. b)~Field intensity for different excitation energies close to the exciton 1s resonance ($1.625\,\si{eV}$ after polaron shift~\cite{DominikSelfenergy}). For $z<0$, the intensity oscillates due to interference effects, and for $z>0$ its value remains constant. c)~Absorption spectra according to Eqs.~(10 - 12) for monolayer MoSe$_2$ embedded in SiO$_2$. The normalization factor is shown in brackets. The values for the radiative dephasing and the homogeneous dephasing are provided in the appendix, table I. The excitonic transition is broadened by the non-Markovian-phonon scattering, which leads also to the formation of a phonon sideband, where the sideband is marked by an arrow. d)~Same plot as (c), but logarithmic scale.}}
    \label{MonolayerSpectra}
\end{figure}

The geometry we investigate is a TMDC monolayer embedded in a structured dielectric environment with refractive index $n(z)$, with increasing complexity, cf. Fig.~\ref{MonolayerSpectra}(a), \ref{Suppressive_Structure}(a), \ref{Enhancing_Structure}(a). To obtain experimental observables such as absorption and reflection, we solve the corresponding wave equation~\cite{Stroucken} for perpendicular propagation in the z-direction:
%\begin{align}
%    \left(\frac{\partial^2}{\partial z^2}-\frac{n^2(z)}{c_0^2}\frac{\partial^2}{\partial t^2}\right)\mathbf{E}(\mathbf{r},t)=\frac{1}{\epsilon_0c_0^2}\frac{\partial^2}{\partial t^2}\mathbf{P}(\mathbf{r},t)
%\end{align}
Here, the macroscopic polarisation $\mathbf{P}$ represents the excitonic transitions of the TMDC and $n(z)$ the refractive index of the structured dielectric surrounding. The extension of the TMDC is small compared to the wavelength of the light momentum with the excitonic transition. The polarisation is approximated within the monolayer as a two dimensional dipole density $\mathbf{P}(\mathbf{r},t)=\delta(z)\mathbf{P}^{2D}(t)$~\cite{Stroucken} at position $z=0$:
\begin{align}
    \left(\frac{\partial^2}{\partial z^2}-\frac{n^2(z)}{c_0^2}\frac{\partial^2}{\partial t^2}\right)\mathbf{E}^{2D}(z,t)=\frac{1}{\epsilon_0c_0^2}\delta(z)\frac{\partial^2}{\partial t^2}\mathbf{P}^{2D}(t)\label{Waveeq}
\end{align}
The optical transitions of the TMDC excitons occur for only one direction of circular polarization $\sigma^-$ (or $\sigma^+$) interaction with the $K$ (or $K'$) valley~\cite{MalteMoSelinewidth}, an effect that is referred to as the circular dicroism~\cite{yao2008valley,wang2018colloquium}. Therefore $\mathbf{P}^{2D}(t)$ is expanded in $\sigma^+,\sigma^-$ emissions with the components $P^{\sigma\pm}$. \changes{Without loss of generality, we only discuss here one polarization component, which we denote as $P=P^{\sigma\pm}$ in the following, as due to symmetry the other circular polarization direction will simply interact with the other valley (K,K') in a completely similar way.} %Hier Malte,Stroucken,etc. zitieren; am besten noch andere Quellen heraussuchen.
The wave equation Eq.~(\ref{Waveeq}) has two solutions.
A forward propagating wave $(f)$ from the left to the right, and a backward propagating wave $(b)$ travelling from the right to the left. For a spatially homogeneous environment $n(z)=n$, cf.~Fig.~\ref{MonolayerSpectra}(a), the solutions for a forward $E^f$ and backward $E^b$ propagating field read~\cite{Stroucken}:
\begin{align}
    E_{T}^{f}\left(t-\frac{nz}{c_0}\right)&=E_{0}^{f}\left(t-\frac{nz}{c_0}\right)-\frac{1}{2n\epsilon_0c_0}\frac{\partial}{\partial t}P\left(t-\frac{nz}{c_0}\right) \label{Transmitted}\\
    E_R^{b}\left(t+\frac{nz}{c_0}\right)&=E_m^{b}\left(t+\frac{nz}{c_0}\right)-\frac{1}{2n\epsilon_0c_0}\frac{\partial}{\partial t}P\left(t+\frac{nz}{c_0}\right) \label{Reflected}
\end{align}
$E_0^{f}$ and $E_m^{b}$ are externally applied fields, and both solutions differ by their signs of the arguments as they form forward or backward propagating solution. Using Eqs.~(\ref{Transmitted}, \ref{Reflected}) both the transmitted field $E_T$ and the reflected field $E_R$ can be determined, once $P$ is known.
\subsection{Excitonic response}
The dipole density $P$ in the wave equation is determined by excitonic transitions and can be derived by calculating the Heisenberg equation of motion for a two band Hamiltonian including the Coulomb and electron-phonon interaction~\cite{MalteMoSelinewidth,DominikSelfenergy,katzer2023impact,Berechnungphi,KiraKochBuch,FrankLengers}. %Malte zitieren, Dominik zitieren, etc zitieren...

In the excitonic basis $P(t)$ reads~\cite{KiraKochBuch}:
\begin{align}
    P(t)=\frac{1}{A}\sum_{q,{\mu}}\left[(\phi_q^\mu)^* \cdot d^{c,\nu} P^*_{\mu,Q}(t)+\phi_q^\mu\cdot (d^{c,\nu})^* P_{\mu,Q}(t)\right],\label{MakroPolMikroPol}
\end{align}
\changes{with the center of mass momentum $\QQ$ and the relative momentum $\qq$ of the exciton. The latter is the in-plane momentum of the excitonic wave function $\phi^\mu_q$ in Fourier space, and $\mu$ the index from the respective exciton Rydberg state. The eigenfunctions and eigenvalues are obtained from solving the Wannier equation~\cite{MalteMoSelinewidth, Florian, DominikSelfenergy, KiraKochBuch,berghauser2014analytical}. $A$~equals the size of the quantization area~\cite{KiraKochBuch} and $d^{c,\nu}$ represents the transition matrix dipole element on the elementary cell between the conduction band $c$ and the valence band $\nu$ of the TMDC at the high symmetry point $K$~\cite{Berechnungphi}.}
This notation includes all excitonic transitions, however, due to the high excitonic binding energy and due to strong Coulomb coupling, in the following section we consider only the spectral range of the $\mu=1s$ excitonic transition.
The first term in this equation is off-resonant for the excitonic transition and can be disregarded in a rotating wave approximation. 
The Bloch equation for the excitonic transition $P_{\mu,Q}(t)$ in the low density limit reads~\cite{DominikSelfenergy, selig2018dark, katzer2023impactnessy}:
%Zitieren von Dominik?
\begin{align}
    &(\hbar\partial_t+iE_\mu(Q))P_{\mu,Q}(t)= \nonumber\\
    &=\frac{i}{A}\sum_q(\phi^{\mu}_q)^*d^{c,\nu}\cdot  E_{total}(t)\delta_{Q,0}+ \nonumber\\ &-i\sum_{\lambda,q',\alpha}g^{\mu\lambda,\alpha}_{q'}(S_{Q+q',q'}^{\lambda,\alpha}(t)+\tilde{S}_{Q+q',-q'}^{\lambda,\alpha}(t)).
    \label{BlochPhonon2}
\end{align}

For a homogeneous structure, the field strength at the TMDC is $E_{total}(t)=E_T^f(t)$ and has to be determined from the solutions of the wave equation Eq.~(\ref{Transmitted}, \ref{Reflected}). \changes{$S_{Q+q',q'}^{\lambda,\alpha}(t)$ represents a phonon assisted exciton transition, with phonon mode $\alpha$ and the transfer of momentum $q'$. It is defined as $S^{\mu,\alpha}_{Q+q',q'}=\langle P_{\mu,Q+q'}b^\alpha_{q'} \rangle$. Here, $b_q^\alpha$ represents the annihilation operator of one phonon in mode $\alpha$. Similarly, $\tilde{S}_{Q+q',-q'}^{\lambda,\alpha}=\langle P_{\mu,Q+q'}b^{\dagger\alpha}_{-q'} \rangle$ describes the emission of a phonon of mode $\alpha$ and transfer of momentum $-q'$.} Their corresponding equations of motion read~\cite{DominikSelfenergy}:
\begin{align}
    &(\hbar\partial_t+iE_\mu(Q+q')+iE_{q'}^{\alpha})S_{Q+q',q'}^{\mu,\alpha}(t)=\nonumber\\
    &=-i\sum_{\lambda} g^{\mu\lambda,\alpha}_{q'}(1+n_{q'}^\alpha)P_{\lambda,Q}(t) \label{S1-2}\\
    &(\hbar\partial_t+iE_\mu(Q+q')-iE_{q'}^{\alpha})\tilde{S}_{Q+q',-q'}^{\mu,\alpha}(t)=\nonumber\\
    &=-i\sum_{\lambda} g^{\mu\lambda,\alpha}_{q'}n_{q'}^\alpha P_{\lambda,Q}(t)\label{S2-2}
\end{align}
\changes{with the exciton-phonon coupling matrix $g^{\mu\lambda,\alpha}_{q'}$, where $\mu$ and $\lambda$ represent excitonic states and $\alpha$ the phonon mode with momentum $q'$ and phonon energy $E_{q'}^\alpha$~\cite{DominikSelfenergy}. } \changes{The phonon occupation number $n_{q'}^\alpha = (\exp(-\frac{E^\alpha_{\qq'}}{k_BT})-1)^{-1}$ is given by the Bose-Einstein-distribution and is thus a temperature dependent quantity. Therefore, by introducing exciton-phonon interaction to our model, the dynamics of the TMDC excitons become temperature dependent~\cite{DominikSelfenergy}}.
For optical transitions only the vanishing center of mass momentum $Q=0$ in the light cone is excited $P_{\mu,Q}\approx P_{\mu}\delta_{Q,0}$~\cite{zhang2015experimental}.
Eqs.~(\ref{Transmitted} - \ref{MakroPolMikroPol}) contain the macroscopic dipole density and the resulting electric fields will be self consistently solved.
In the frequency domain, Eq.~(\ref{BlochPhonon2}-\ref{S2-2}) can be solved, where the exciton-phonon scattering is represented by the phonon-induced self energy $\Sigma(\omega,T)$~\cite{DominikSelfenergy}: 
\begin{align}
P_{1s}(\omega)&=\frac{\frac{1}{A}\sum_{q'}(\phi^{{1s}}_q)^*d^{c,\nu}E^f_{0}(\omega)}{-\hbar\omega-i\hbar\gamma_r+E_{1s}-\Sigma(\omega,T)} \label{PfürSigma}
\end{align}
This self energy $\Sigma(\omega,T)$ introduces a linewidth broadening and polaronic frequency shift~\cite{DominikSelfenergy,FrankLengers}. 
The radiative dephasing in Eq.~(\ref{PfürSigma}) is self consistently calculated by inserting Eq.~(\ref{MakroPolMikroPol}) in Eq.~(\ref{BlochPhonon2}) and reads:
\begin{align}
\hbar\gamma_r=\frac{\omega}{2\epsilon_0 n c_0}\frac{1}{A^2}\bigg|\sum_q \phi_q^{1s}\cdot (d^{c,\nu})^*\bigg|^2, \label{RadDephMonolayer}
\end{align}
For $n(z)=n=const.$ the transmission and reflection spectra can be analytically determined by the ratio of the field intensity with respect to the incident field intensity~\cite{KiraKochBuch, AndreasBuch}:
\begin{align}
    &T(\omega)=\frac{|E_T|^2}{|E_0|^2}=\bigg|1+\frac{i\hbar\gamma_r}{-\hbar\omega+E_{1s}-i\hbar\gamma_r-\Sigma(\omega,T)}\bigg|^2 \label{MonoT}\\
    &R(\omega)=\frac{|E_R|^2}{|E_0|^2}=\bigg|\frac{i\hbar\gamma_r}{-\hbar\omega+E_{1s}-i\hbar\gamma_r-\Sigma(\omega,T)}\bigg|^2 \label{MonoR}\\
    &\alpha(\omega)=1-T(\omega)-R(\omega)\label{Monoa}
\end{align}
The absorption spectrum for MoSe$_2$ is shown in Fig.~\ref{MonolayerSpectra}~b,c), where next to the main resonant transition at $\hbar\omega\approx1.64\,\si{eV}$ a phonon sideband can be observed at $\hbar\omega\approx1.75\,\si{eV}$. \changes{Here, the initially pure single excitonic transition at $E_{1s}=1.68\,\si{eV}$ was considered. This main excitonic transition is red shifted by the non-Markovian self-energy. Moreover, a sidepeak occurs due to phonon assisted absorption generating momentum dark excitons, where the exciton-phonon coupling provides the momentum for an indirect coupling to the light field. In the absorption spectrum, this leads to a modification of the single excitonic Lorentzian lineshape by phonon sidebands. This effect is seen in experiments~\cite{DominikVerspannungen}, and also in photoluminescence~\cite{brem2020phonon}. It was explained with the mentioned non-Markovian theory by our group in Ref.~\cite{DominikSelfenergy}, which was later expanded in Refs.~\cite{shree2018observation,FrankLengers}. We use here material realistic values for MoSe$_2$, to correctly predict the lineshape as it is expected in a respective experiment at room temperature~\cite{DominikSelfenergy}. All material specific parameters used in Eqs. (\ref{S1-2} - \ref{Monoa}) to calculate the absorption spectrum were obtained from \textit{ab initio} literature and are listed in App.~A.}
\subsection{Dielectric structure}\label{Sec:Dielec}
The effect of time delayed feedback control is introduced through a dielectric structure shown in Figs.~~\ref{Suppressive_Structure}(a),~\ref{Enhancing_Structure}(a). For this purpose the TMDC is considered to be encapsulated within a dielectric of refractive index $n(z)$ with the top layer width $\Delta_2$ and bottom layer width $\Delta_1$ and a mirror at position $\Delta_1$ reflecting the transmitted field from the TMDC.
This way the optical response of the TMDC $P(t)$ is influenced by its time delayed response $P(t-\tau_i)$ with a time delay $\tau_i=\frac{2n\Delta_i}{c_0}$ \changes{introduced by} the optical path time between TMDC and mirror/interface~\cite{Florian,carmele2013single,finsterholzl2020nonequilibrium,barkemeyer2020revisiting}. 
%The total field strength $E_{total}$ then includes the original response of the TMDC and a time delayed feedback response. 
By positioning the mirror at varying distances $\Delta_1$ behind the TMDC this time delay can be varied. We calculate the fields in the time domain similarly to the previous section, starting with the relevant boundary conditions for the electric field. We identify the following five equations from the boundary conditions between the different dielectrics (air/vacuum-SiO$_2$) and the reflection at the mirror position as well as the transmission through the TMDC. These equations depend on the width of the top layer $\Delta_2$, as well as the mirror's position $\Delta_1$:
\begin{align}
    E_{T1}\left(t+\frac{n\Delta_2}{c_0}\right)&=t^+E_{0}\left(t+\frac{n\Delta_2}{c_0}\right)+r^+E_{R1}\left(t-\frac{n\Delta_2}{c_0}\right) \label{ForwardTopLayer}\\
    E_{R}\left(t-\frac{n\Delta_2}{c_0}\right)&=t^-E_{R1}\left(t-\frac{n\Delta_2}{c_0}\right)+r^-E_{0}\left(t+\frac{n\Delta_2}{c_0}\right) \label{BackwardTopLayer}\\
    E_{T}\left(t\right)&=E_{T1}\left(t\right)-\frac{1}{2n\epsilon_0c_0}\frac{\partial}{\partial t}P(t) \\
    E_T\left(t-\frac{n\Delta_1}{c_0}\right)&=-E_B\left(t+\frac{n\Delta_1}{c_0}\right) \label{Mirrorequation}\\
    E_{R1}(t)&=E_B(t)-\frac{1}{2n\epsilon_0c_0}\frac{\partial}{\partial t}P(t)
\end{align}
The Fresnel coefficients for perpendicular incident fields read $t^+=\frac{2n_{vac}}{n_{vac}+n_{SiO_2}}$, $t^-=\frac{2n_{SiO_2}}{n_{vac}+n_{SiO_2}}$, $r^+=\frac{n_{SiO_2}-n_{vac}}{n_{vac}+n_{SiO_2}}$ and $r^-=\frac{n_{vac}-n_{SiO_2}}{n_{vac}+n_{SiO_2}}$~\cite{Toulouse}. %Hier Toulouse und eventuell andere Quellen zitieren.
Eqs.~(\ref{ForwardTopLayer}, \ref{BackwardTopLayer}) represent the transition from vacuum/air to SiO$_2$. Eq.~(\ref{Mirrorequation}) represents the condition that the field strength at the surface of the mirror vanishes, i.e.~the existence of a fully reflecting mirror.
In order to calculate the optical spectra of the TMDC, the reflected field strength $E_R$ and the total field strength $E_{total}=E_{R1}(z=0)+E_{T1}(z=0)=E_{B}(z=0)+E_T(z=0)$ have to be determined solely depending on the incident field strength $E_0$ and $P$. By transforming all equations into Fourier space, the resulting equations form a linear system of equations, similarly to the transfer matrix method~\cite{Toulouse}. %Hier Toulouse zitieren
\changes{The details of the calculation can be found in the Appendices~\ref{app:fourier}, \ref{app:feedback}.}
We can neglect the transmission through the mirror, thus only the reflected intensity and absorption have to be calculated:
\begin{align}
    &R(\omega)\\
    &=\bigg|A_{R}+\frac{i\hbar\gamma_r(A_{T1}+A_{R1})B_R}{-\hbar\omega+E_{1s}+\delta_r(\tau_1,\tau_2)-i\hbar\gamma_r(\tau_1,\tau_2)-\Sigma(\omega,T)}\bigg|^2\nonumber\\
    &\alpha(\omega)=1-R(\omega)\label{eq:absorption}
\end{align}
with the phonon induced self energy $\Sigma(\omega)$ introduced in Eq.~(\ref{PfürSigma}) and the time delay \changes{induced} radiative dephasing 
\begin{align}
    &\hbar\gamma_r(\tau_1,\tau_2)=\hbar\gamma_r\,\si{Im}[B_{T1}+B_{R1}] \label{RadDephcontrolled}\\
    &\hbar\gamma_r=\frac{\omega}{2\epsilon_0 n c_0}\frac{1}{A^2}\bigg|\sum_q \phi_q^{1s}\cdot (d^{c,\nu})^*\bigg|^2 \label{RadDephMono}
\end{align}
and a radiative frequency shift are introduced
\begin{align}
    \delta_r(\tau_1,\tau_2)=\hbar\gamma_r\,\si{Re}[B_{T1}+B_{R1}]\label{Energyshift}
\end{align}
 The complex coefficients $A_i,\,B_i$, cf. App.~\ref{app:feedback}, Eq.~(\ref{Coefficients_inhomogeneous_Structure}), represent geometry induced phase factors and influence the feedback controlled radiative dephasing $\hbar\gamma_r(\tau_1,\tau_2)$, which differs clearly from the radiative dephasing for the homogeneous structure cf. Eqs. (\ref{RadDephMonolayer}, \ref{RadDephMono}). %Hier Formel für Monolayer referencen
The energy shift (Eq.~(\ref{Energyshift})) induced by feedback induced interference has been theoretically predicted~\cite{Florian} %Hier Florians Paper zitieren
as a polaritonic frequency shift, which changes the position of the resonant transition. In this work we want to focus on the time delayed radiative dephasing $\hbar\gamma_r(\tau_1,\tau_2)$, which is now a function of the mirrors position and may be increased or reduced compared to the radiative dephasing of the monolayer $\hbar\gamma_r$. \changes{The extension to multi-layered structures is straightforward~\cite{Stroucken} and should, for a careful design, conserve the interference effects described here for a single layer.}
A discussion of the absorption spectra, Eq.~(\ref{eq:absorption}), is provided in Sec.~\ref{Sec:Disc}.
\section{Discussion}\label{Sec:Disc}
\begin{figure}[t!]
    \centering
    \includegraphics[width=0.95\linewidth]{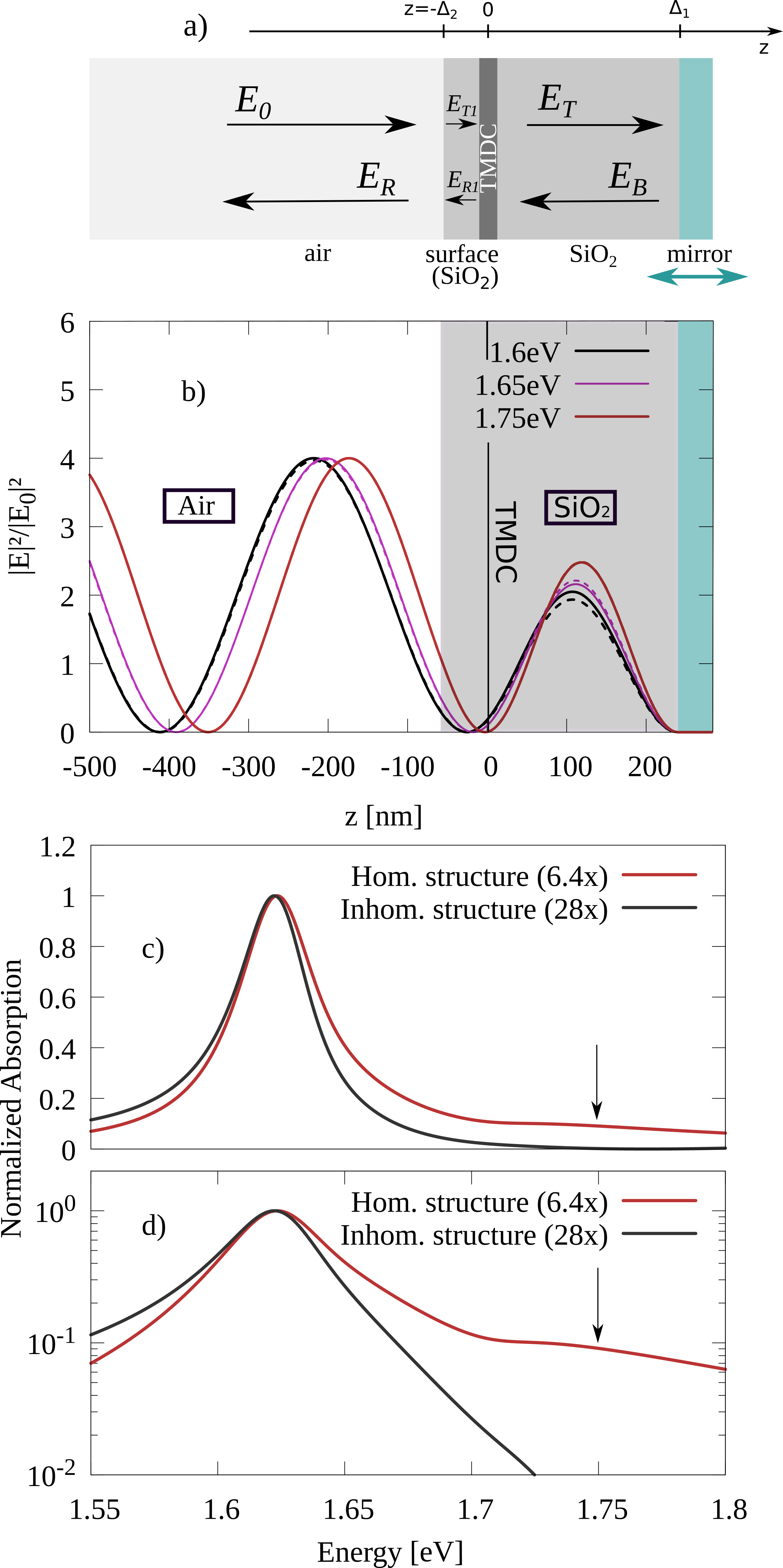}
    \caption{\changes{a) Sketch of the geometry of the inhomogeneous structure. The turquoise arrow indicates the tunability between constructive or destructive interference by varying the thickness of the dielectric material as a function of the mirror position. b)~Field intensity profile, without the TMDC (solid lines) and with TMDC included (dashed lines). The area in gray represents the surrounding dielectric medium SiO$_2$, with the TMDC at position $z=0$. The structure parameters (cp. (a)) are $\Delta_1=240\,\si{nm}$ and $\Delta_2=60\,\si{nm}$. The electric field intensity at the TMDC ($z=0$) vanishes for $\hbar\omega=1.75\,\si{eV}$, i.e. at the spectral position of the phonon sideband.  c)~Comparison of the absorption spectra for the geometry of Figs.~1(a) and 2(a): Suppression of the phonon sideband absorption $\alpha(\omega)$ due to negligible optical interaction for its spectral position, and a general reduction of $\alpha(\omega)$ occurs. d)~Same absorption in a logarithmic scale.}}
    \label{Suppressive_Structure}
\end{figure}
\begin{figure}[t!]
    \centering
    \includegraphics[width=0.95\linewidth]{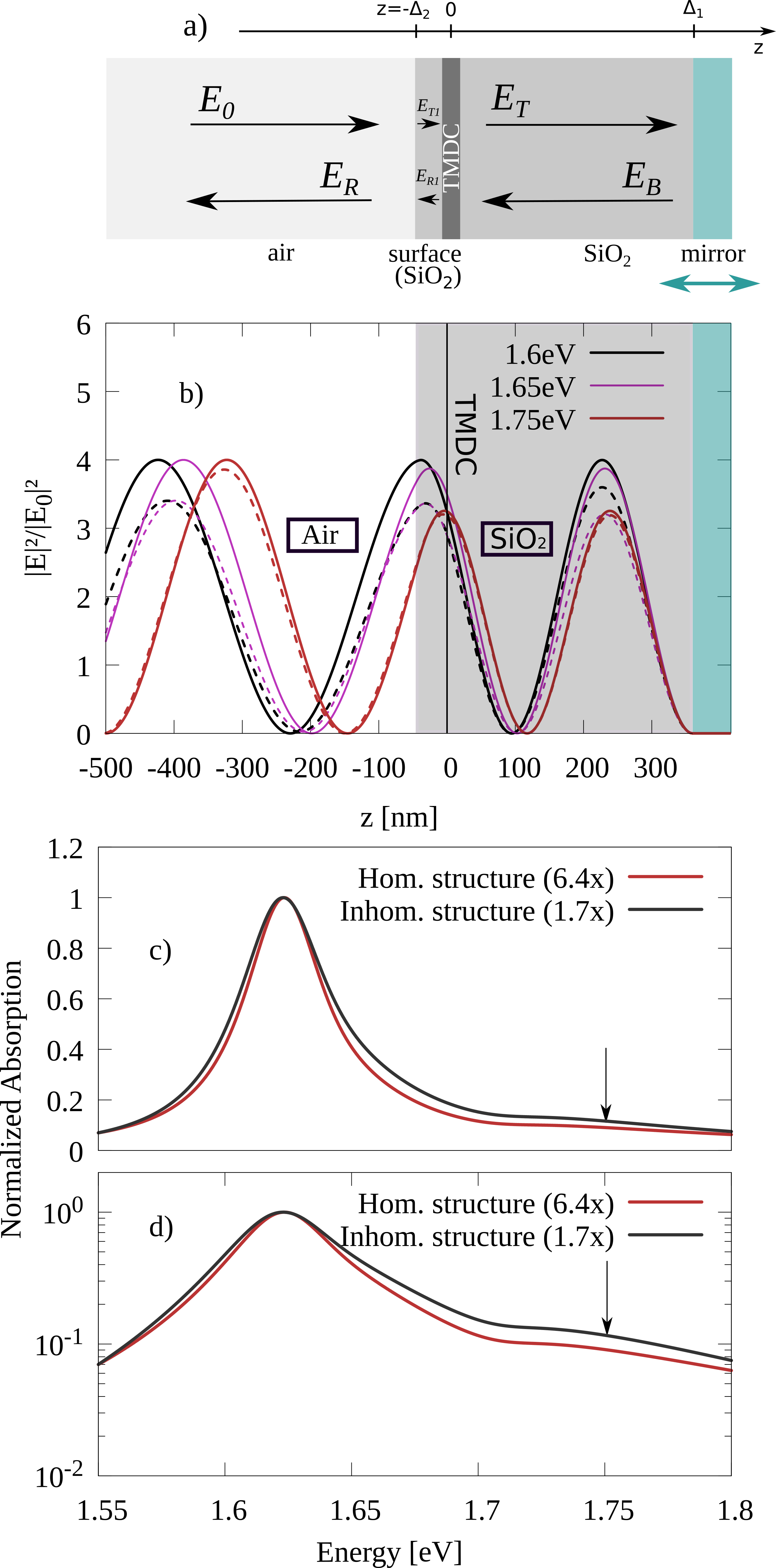}
    \caption{As in Fig.~\ref{Suppressive_Structure}, but the parameters are $\Delta_1=360\,\si{nm}$ and $\Delta_2=40\,\si{nm}$. \changes{The increased separation $\Delta_1$ shown in a) yields an increase of the field intensity, b), at the position of the TMDC ($z=0$) through constructive interference compared to the case of a monolayer.} This has the effect of increasing the radiative coupling strength, which increases the overall absorption, c),(d) and amplifies the phonon sideband at $\hbar\omega\approx1.75\,\si{eV}$.}
    \label{Enhancing_Structure}
\end{figure}

In the previous section the models of the linear absorption spectra involving the interplay of non-Markovian exciton-phonon scattering at the temperature $T=300\,\si{K}$ and mirror induced time delayed feedback were derived. The corresponding radiative dephasing develops according to Eq.~(\ref{RadDephcontrolled}), where the non-Markovian exciton-phonon scattering introduces phonon sidebands. In this section, two possible setups,\changes{ Figs.~\ref{Suppressive_Structure}(a) and \ref{Enhancing_Structure}(a), }will be discussed, where the absorption is influenced by the feedback, i.e.~either enhanced through constructive interference at the TMDC or decreased via destructive interference. To illustrate this scheme, we use the material sytem of MoSe$_2$, where phonon sidebands are already observed \changes{experimentally}, due to strong exciton - optical phonon coupling~\cite{DominikSelfenergy,FrankLengers}: \changes{In this material,} in addition to the main resonant transition for a homogeneous dielectric environment, a pronounced phonon sideband can be observed at $\hbar\omega\approx1.75\,\si{eV}$ in Fig.~\ref{MonolayerSpectra}(c),(d), thus reproducing the results of Ref.~\cite{DominikSelfenergy}. The enhancement or suppression of this sideband will be the focus of our study. The mirror \changes{introducing the optical feedback} is now positioned such that phonon sidebands are optically suppressed, by tuning the structures parameters $\Delta_1, \Delta_2$, such that the radiative dephasing vanishes at this position.
In Fig.~\ref{Suppressive_Structure}(a), one possible structure is shown, where $\Delta_1=240\,\si{nm}$ for a top layer width of $\Delta_2=60\,\si{nm}$. Here, the electric field intensity profile in Fig.~\ref{Suppressive_Structure}\text{b)} shows that the electric field destructively interferes at the position of the TMDC, which suppresses optical interaction at the frequency of the phonon sideband. In Fig.~\ref{Suppressive_Structure}(c),(d) the calculated absorption spectrum is shown. It can be seen that, in principle, the whole spectrum is influenced by the \changes{geometry induced} decrease of absorption, cf. the \changes{normalization} factors in Fig.~\ref{Suppressive_Structure}(c). The feedback causes the absorption \changes{at the energetic position of the sideband} to vanish completely, due to $\hbar\gamma_r(\tau_1,\tau_2,1.75\,\si{eV})\approx 0$. 
This setup shows that besides the possibility to control the radiative dephasing, destructive interference from a mirror can be used to suppress phonon sidebands from appearing within the optical absorption spectra for TMDCs.
In a second setup, the value of the radiative dephasing is increased by constructively interfering the field at the position of the TMDC. For $\Delta_1=360\,\si{nm}$ and a top layer width of $\Delta_2=40\,\si{nm}$, the mode profile and optical spectra are shown in Fig.~\ref{Enhancing_Structure}. From \changes{the comparison of the mode profile in Figs.~\ref{MonolayerSpectra}(b) and} \ref{Enhancing_Structure}(b), it can be concluded that the field intensity is enhanced by three times of the incident field intensity. The constructive interference increases the radiative dephasing and yields an increase of the overall absorption shown in Fig.~\ref{Enhancing_Structure}\text{c)}. Thus the presented structure may be applied to \changes{increase} optical absorption, by improving the ratio of radiative dephasing to the phonon induced broadening, which could help pronounce and identify phonon sidebands of TMDCs.

\section{Conclusion}
The \changes{optical absorption} for a monolayer MoSe$_2$ at $T=300\,\si{K}$ was calculated with respect to the interplay of non-Markovian exciton-phonon scattering and feedback induced radiative dephasing. Two possible \changes{geometries} for MoSe$_2$ were presented, one where the absorption of a phonon sideband was suppressed through destructive interference at the TMDC, and another structure where the \changes{sideband was enhanced} through constructive interference.~\changes{An experimental validation of the predicted exciton-phonon interaction control is not restricted to a single TMDC layer, where the radiative decay can be suppressed or amplified with feedback~\cite{Toulouse}, and recently also the experimental control of bright and dark exciton splitting was reported~\cite{ren2023control}. It could be extended to multi-layer samples~\cite{Stroucken} and also be used to enhance the coherence and quantum correlations of light by feedback in cavities~\cite{carmele2013single} by suppressing the exciton-phonon coupling. For organic semiconductors, typically involving a series of phonon sidebands~\cite{schweicher2019chasing}, and for quantum dots~\cite{favero2003acoustic,wigger2020acoustic}, the control of phonon sidebands are also in reach. Besides, since similar phonon replica are also known to occur for cavity exciton-polariton setups~\cite{lengers2021phonon}, it would in principle be intriguing to see whether additional mirrors could be used to apply the suggested suppression scheme also in related systems. This might however not be very straightforward, as the \textit{feedback} mirror would introduce a trade-off between strong coupling and the control of phonon sidebands. Also an application of the proposed setup to localized plasmons, e.g., in hybrid systems with metal nanoparticles, where peaked resonances occur in respective spectra~\cite{salzwedel2023spatial,greten2023dipolar}, could be a subject to control.}

\section{Acknowledgments}
We thank Florian Katsch, Malte Selig, Dominik Christiansen, Kisa Barkemeyer and Fares Schulz for helpful discussions and advice. We gratefully acknowledge funding by the Deutsche Forschungsgemeinschaft (DFG) through SFB 910 (Project No.~163436311) (T. T.) and through SFB 951 (Project No.~182087777) (M. K., A. K.).

\appendix
% \section{Appendix}
\section{Parameters}\label{Sec:Append}
\begin{table}[h!]
    \centering
    \begin{tabular}{|l|c|}
    \hline
        $\hbar\gamma_{r_0}$ (MoSe$_2)$ & 2.2\,\si{meV}~\cite{MalteMoSelinewidth} \\
         \hline
        $\hbar\gamma_0$ (MoSe$_2$, $300\,\si{K}$) & 18\,\si{meV}~\cite{MalteMoSelinewidth} \\
         \hline
        %$n_{hBN}$~\cite{Florian},~\cite{RefrhBN} & 2.12 (calculated)\\
        %\hline
        $n_{SiO_2}$ & 1.46 ($1.6\,\si{eV}$)~\cite{gao2013refractive}\\
        \hline
        $E_{1s}$ (MoSe$_2$) & 1.68\,\si{eV}~\cite{DominikVerspannungen}\\
        \hline
        $g_{q}^\alpha$ (MoSe$_2$) & 50\,\si{meV}~\cite{DominikSelfenergy,selig2020suppression}\\
        \hline
        $E_{q}^\alpha$ (MoSe$_2$) & 37\,\si{meV}~\cite{DominikSelfenergy,selig2020suppression}\\
        \hline
    \end{tabular}
    \caption{Material Parameters of the investigated monolayer MoSe$_2$.}
    \label{tab:my_label}
\end{table}

\section{Details on the Fourier transform}\label{app:fourier}
In Sec.~\ref{Sec:Theo}, the properties of the Fouriertransform were applied, in order to determine the electric field strength and to solve the Bloch equation. The applied properties will be derived in this section. First we identify the definition of the Fouriertransform, which transforms a function into Fourier space.
\begin{align}
    f(\omega):=\int dt f(t)e^{i\omega t}
\end{align}
Certain properties may be derived for derivatives or arguments of $f(t)$ %Hier das Ingenieur oder Mathebuch zitieren:
\begin{align}
    \int dt f(t-\tau)e^{i\omega t}=\int dsf(s)e^{i\omega (s+\tau)}=e^{i\omega\tau}f(\omega)\\
    \int dt \partial_t(f(t))e^{i\omega t}=-\int dt e^{i\omega t}(i\omega)f(t)=-i\omega f(\omega)
\end{align}
The two properties may both be applied as is in the case for $\partial_t(P_\mu(t-\tau))$, and were used to calculate the optical spectra shown in Sec.~\ref{Sec:Theo}.
\\
\\
\section{Details on the feedback control}\label{app:feedback}
For the inhomogeneous structure, cf.~Figs.~\ref{Suppressive_Structure}(a),~\ref{Enhancing_Structure}(a), the field strengths as a function of the incident field strength are represented in the frequency domain through the following system of linear equations:
\begin{align}
\begin{pmatrix}
1 & 0 & 0 & 0 & 0 & 0 & 0 \\
t^+ & 0 & 0 & 0 & r^+g_2^+ & 0 & 0 \\
0 & 1 & 0 & 0 & 0 & 0 & i\kappa \\
r^-g_2^- & 0 & 0 & 0 & t^- & 0 & 0  \\
0 & 0 & 0 & 0 & 0 & 1 & i\kappa\\
0 & 0 & -g_1^+ & 0 & 0 & 0 & 0 \\
0 & 0 & 0 & 0 & 0 & 0 & 1 \\
\end{pmatrix}
\begin{pmatrix}
E_0\\
E_{T1}\\
E_{T}\\
E_R\\
E_{R1}\\
E_{B}\\
P
\end{pmatrix}
=
\begin{pmatrix}
E_0\\
E_{T1}\\
E_{T}\\
E_R\\
E_{R1}\\
E_{B}\\
P
\end{pmatrix},
\end{align}
%Hier die Matrix mit den Gleichungen 
where the delay times $\tau_1=\frac{2n\Delta_1}{c_0}$ and $\tau_2=\frac{2n\Delta_2}{c_0}$ were introduced, which represent the optical path time between the TMDC and the mirror and the optical path time between the top layer and the TMDC. The optical phase shift can be calculated through $g_j^+=\si{Exp}(i\omega\tau_j)$, which influences the field strength when interfering at a surface or TMDC. The solution depending on the incident field $E_0$ and $P$ reads
%Hier der Vektor im Fourierraum
\begin{align}
\begin{pmatrix}
E_0\\
E_{T1}\\
E_{T}\\
E_R\\
E_{R1}\\
E_{B}\\
P
\end{pmatrix}
=
\begin{pmatrix}
1\\
A_{T1}\\
A_{T}\\
A_{R}\\
A_{R1}\\
A_{B}\\
0
\end{pmatrix}E_0+i\kappa
\begin{pmatrix}
0\\
B_{T1}\\
B_{T}\\
B_{R}\\
B_{R1}\\
B_{B}\\
1
\end{pmatrix}P,
\label{Coefficients_inhomogeneous_Structure}
\end{align}
where
\begin{align}
\begin{pmatrix}
A_{T1}\\
A_{T}\\
A_{R}\\
A_{R1}\\
A_{B}
\end{pmatrix}
=
\begin{pmatrix}
\frac{t^+}{1+g_1^+g_2^+r^+}\\
\frac{t^+}{1+g_1^+g_2^+r^+}\\
\frac{r^-}{g_2^+}-\frac{g_1^+t^+t^-}{1+g_1^+g_2^+r^+}\\
\frac{-g_1^+t^+}{1+g_1^+g_2^+r^+}\\
\frac{-g_1^+t^+}{1+g_1^+g_2^+r^+}
\end{pmatrix}\\
\begin{pmatrix}
B_{T1}\\
B_{T}\\
B_{R}\\
B_{R1}\\
B_{B}
\end{pmatrix}
=
\begin{pmatrix}
\frac{g_2^+r^++g_1^+g_2^+(r^+)^2}{1+g_1^+g_2^+r^+}\\
\frac{1+g_2^+r^+}{1+g_1^+g_2^+r^+}\\
\frac{-g_1^+t^-+t^-}{1+g_1^+g_2^+r^+}\\
\frac{1-g_1^+}{1+g_1^+g_2^+r^+}\\
\frac{-g_1^+g_2^+r^+-g_1^+}{1+g_1^+g_2^+r^+}
\end{pmatrix},
\end{align}
which allows us to insert the relevant fields into the Bloch equations and calculate the excitonic transition:
\begin{align}
P_{{1s}}(\omega)=\frac{\frac{1}{A}\sum_q(\phi_{{1s}}^q)^*d^{c,\nu}(A_{T1}+A_{R1})E_0(\omega)}{-\hbar\omega+E_{1s}+\delta_r(\tau_1,\tau_2)-i\hbar\gamma_r(\tau_1,\tau_2)-\Sigma(\omega,T)}, 
\end{align}
where the feedback controlled radiative dephasing
\begin{align}
    &\hbar\gamma_r(\tau_1,\tau_2)=\hbar\gamma_r\cdot\si{Im}[B_{T1}+B_{R1}],
    \end{align}
    \begin{align}
    &\hbar\gamma_r=\frac{\omega}{2\epsilon_0 n c_0}\frac{1}{A^2}\bigg|\sum_q \phi_q^{1s}\cdot (d^{c,\nu})^*\bigg|^2,
\end{align}
and a radiative frequency shift are introduced
\begin{align}
    \delta_r(\tau_1,\tau_2)=\hbar\gamma_r\cdot\si{Re}[B_{T1}+B_{R1}].
\end{align}
This energy shift induced by feedback induced interference has been theoretically predicted~\cite{Florian} %Hier Florians Paper zitieren
as a polaritonic frequency shift, which changes the position of the resonant transition. In this work we want to focus on the radiative dephasing, which is now a function of the mirrors position and may be increased or reduced compared to the radiative dephasing of the monolayer $\hbar\gamma_r$. The complex coefficients $A_i,\,B_i$ represent the phase factors and influence the feedback controlled radiative dephasing $\hbar\gamma_r(\tau_1,\tau_2)$, which differs from the radiative dephasing for the homogeneous structure cf.~Eqs.~(\ref{RadDephMonolayer}, \ref{RadDephMono}). %Hier Formel für Monolayer referencen
The mechanism is also dependent on the frequency $\omega$. One possible classical interpretation might be the frequency dependence of the interference pattern, which is shown in the mode profiles cf.~Figs.~\ref{Suppressive_Structure}(b), \ref{Enhancing_Structure}(b).

\bibliography{Bib}
\end{document}